\documentclass[a4paper,UKenglish,creativecommons,nocommercials]{eptcs}

\usepackage[utf8]{inputenc}

\usepackage{url}

\usepackage{xspace}

\usepackage{xcolor}

\usepackage{framed}

\usepackage[inline]{enumitem}

\usepackage{verbatim}

\newcommand{\remove}[1]{}

\usepackage{amsmath}
\usepackage{amsfonts,amssymb}

\usepackage[standard,thmmarks]{ntheorem}

\newcommand{\NN}{\mathbb{N}}

\newcommand{\esim}{\sim_{\!\scriptscriptstyle E}}
\newcommand{\econg}{\simeq_{\!\scriptscriptstyle E}}
\newcommand{\eto}{\to_E}
\newcommand{\sigcong}{\simeq_{\sigma}}
\newcommand{\sigto}{\to_{\sigma}}

\newcommand{\bto}{\to_\beta}
\newcommand{\btod}{\to_{\beta_d}}
\newcommand{\gto}{\to_g}
\newcommand{\lto}{\multimap}
\newcommand{\ltob}{\multimap_\beta}
\newcommand{\ato}{\to_a}
\newcommand{\ltoh}{\lto_h}
\newcommand{\toh}{\to_h}
\newcommand{\beq}{=_{\beta}}

\newcommand{\buco}{\square}
\newcommand{\SV}{\mathsf{SV}}

\newcommand{\FV}{\mathsf{FV}}

\newcommand{\numx}[1]{\texttt{\#}_{#1}}
\newcommand{\nlam}{\numx{\lambda}}
\newcommand{\narg}{\numx{@}}
\newcommand{\nprim}{\numx{p}}
\newcommand{\mymathopstyle}[1]{\mathsf{#1}}

\DeclareMathOperator{\ar}{\mymathopstyle{arg}}

\DeclareMathOperator{\hv}{\mymathopstyle{hv}}

\newcommand{\spo}{\prec}

\newcommand{\imp}{\Rightarrow}
\newcommand{\dom}[1]{[#1]}
\newcommand{\sfun}[1]{[#1]}

\newcommand{\Lambdat}{\Lambda^\to}

\title{%
  Linear $\beta$-reduction\thanks{%
     Partially  supported by the Project ELICA (ref.~ANR-14-CE25-0005), of
     the ANR program ``Fondements du numérique (DS0705) 2014''.
  }
}%

\author{%
   Stefano Guerrini\\%
  LIPN, Institut Galilée, Université Paris Nord 13, Sorbonne Paris Cité\\
  \texttt{stefano.guerrini@univ-paris13.fr}
}

\newcommand{\authorrunning}{S. Guerrini} 

\newcommand{\titlerunning}{Linear $\beta$-reduction}

\begin{document}

\maketitle
                       

\begin{abstract}
  Linear head reduction is a key tool for the analysis of reduction
  machines for $\lambda$-calculus and for game semantics. Its
  definition requires a notion of redex at a distance named primary
  redex in the literature. Nevertheless, a clear and complete
  syntactic analysis of this rule is missing. We present here a
  general notion of $\beta$-reduction at a distance and of linear
  reduction (i.e., not restricted to the head variable), and we
  analyse their relations and properties. This analysis rests on a
  variant of the so-called $\sigma$-equivalence that is more suitable for
  the analysis of reduction machines, since the position along the
  spine of primary redexes is not permuted. We finally show that, in
  the simply typed case, the proof of strong normalisation of linear
  reduction can be obtained by a trivial tuning of Gandy's proof for
  strong normalisation of $\beta$-reduction.
\end{abstract}

\section{Introduction}
\label{sec:introduction}

Linear head reduction is a key tool for the analysis of reduction
machines for $\lambda$-calculus and for game semantics. A detailed
analysis of it, and more generally of a notion of reduction at a
distance, has been given by
Accattoli~\cite{accattoli:LIPIcs:2013:4051} in terms of proof nets and
explicit substitutions.  Linear head reduction is usually presented in
terms of the so-called $\sigma$-equivalence introduced by Regnier
in~\cite{Reg:EquivLamTermes:94}. In the following, we introduce a
variant of the $\sigma$-equivalence, which has the main advantage of
leaving unchanged the order of primary redexes (a notion of
$\beta$-redex that will be discussed later).  Such a new equivalence
is more suitable for the analysis of abstract reduction machines based
on linear head reduction, as for instance Danos and Regnier's Pointer
Abstract Machine (PAM)~\cite{DR:HeadLinRed:04}, which has been
analysed in detail by the author and Pellitta
in~\cite{GP:DissPAM:16}. Indeed, most of the material that we shall
present in this paper has been developed for formalising the results
in~\cite{GP:DissPAM:16}.

The key tool of our approach is a notion of context which is indeed an
implicit representation of environments mapping variables to their
values. By means of these contexts, one can define a $\beta$-reduction
at a distance and its linearised version. Both of these reduction
rules preserve $\beta$-equivalence, and both of them are strongly
normalising in the case of simply typed $\lambda$-calculus. However,
the proof of strong normalisation is not at all evident. In fact,
linear reduction does not erase any term, it just replaces one of the
occurrences of a variable with a (larger) $\lambda$-term; in other
words, the size of the reducing term always increases along the
reduction. Surprisingly, this apparent difficulty can be trivially
overcome by a small tuning of Gandy's proof for strong normalisation
of $\beta$-reduction~\cite{Gan:ProofsSN:80}. Just by changing a detail
in the interpretation of variable occurrences---it suffices to
increase by 1 their measure---we can adapt the measure used in Gandy's
proof to the case of linear reduction. Moreover, the new measure
obtained in this way simultaneously proves strong normalisation of
$\beta$-reduction and of its linearised version.

As already remarked, linear reduction has been studied in detail by
Accattoli~\cite{accattoli:LIPIcs:2013:4051} by means of linear logic
proof nets. Such an approach has been inspired by the structural calculus
introduced by Accattoli and Kesner~\cite{accattoli_structural_2010}, a
calculus with explicit substitutions and reduction rules at a
distance. In the present paper, our goal is to analyse linear
reduction directly on $\lambda$-calculus, without introducing explicit
substitutions or without going down to the low level analysis of
reduction that can be achieved by means of proof nets. As we shall see
later, such a goal is achieved by introducing a variant of
$\sigma$-equivalence, named E-equivalence, which is more suitable for
investigating reduction machines based on pointers, as for instance
the PAM. Moreover, the proof of strong normalisation that we shall
give is much simpler then the one based on reducibility
candidates required in the case of proof nets. Recently, Pedrot and
Saurin~\cite{pedrot_classical_2016} have proposed a call-by-need
variant of $\lambda\mu$-calculus defined in terms of a notion of
closure contexts. Such closure contexts correspond to the E-contexts
introduced in the following by Definition~\ref{def-E-context}, but
extended to $\lambda\mu$-calculus too. We remark that most of the
material that we shall present below is a by-product of the study of
the PAM started in~\cite{GP:DissPAM:16}, and that one of the further
developments mentioned in~\cite{GP:DissPAM:16} is the extension of the
PAM to the $\lambda\mu$-calculus; Pedrot and Saurin's call-by-need
closure contexts seem to be the right tool for formalising such an
extension.

\section{Preliminaries}
\label{sec:preliminaries}

The set of the $\lambda$-terms $\Lambda$ is defined by the abstract
grammar $s,t ::= x \mid \lambda x.t \mid st$, where $x\in\mathcal{V}$
and $\lambda x$ is a binder for the variable $x$.  The set of the free
variables of a term $t$ is denoted by $\FV(t)$.  The key computational
step of $\lambda$-calculus is $\beta$-contraction
$(\lambda x. t) s \bto t\{s/x\}$, where $t\{s/x\}$ denotes that every
free occurrence of the variable $x$ in $t$ is replaced by $s$,
provided that such a replacement does not cause any name clash of some
free variable of $s$; otherwise, if this is not the case, one has to
preliminarily apply a suitable sequence of variable renamings, or
$\alpha$-rules, to $t$. (The $\alpha$-congruence is the least
congruence induced by the $\alpha$-rule
$\lambda x.t = \lambda y.t\{y/x\}$, in which $y$ replaces all the free
occurrences of $x$ in $t$ and $y$ does not occur in $t$.) As
usual, $\bto^*$ denotes the reflexive and transitive closure of the
binary relation defined by the $\beta$-rule, and $\beq$ denotes the
corresponding equivalence (closing by symmetry also). Such notations
will extend to the other rewriting rules that we shall see in the
paper.

In order to avoid the bureaucratic problems connected to
$\alpha$-congruence, we can assume to work modulo it, and
that all the bound variables in the terms that we shall consider have
the \emph{distinct names property} (sometimes referred to as
Baredrengt variable names convention). A term has the distinct names
property if no free variable in it has the same name of a bound
variable, and all the bound variables have distinct names. Remarkably,
for every $\lambda$-term there is an $\alpha$-congruent one which has
the distinct name property.  In this way, no name clash can arise by
replacing $s$ for $x$ in $t$ in the $\beta$-reduction of
$(\lambda x.t)s$.  However, even if correct, the resulting term
$t\{s/x\}$ might not have the above distinct names property. In order
to guarantee that $t\{s/x\}$ preserves the distinct names property of
$(\lambda x. t) s$, we can assume to replace each occurrence of $x$
with a \emph{fresh} copy of $s$, in which every bound variable has a
fresh name which has not been already used in the term or in another
copy of $s$.

In the simply typed $\lambda$-calculus, every term has a type. The set
of types is given by the abstract grammar
$\tau,\sigma::= o \mid \tau \to \sigma$, where the constant $o$ is the
unique \emph{base type} and any type $\tau\to\sigma$ is said a
\emph{functional type}.  The set $\Lambdat$ of the simply typed terms
is the subset of $\Lambda$ whose terms respect the following
typing rules: 
\begin{enumerate*}[label=(\emph{\roman*})]
\item each variable $x$ has a given type $\tau$;
\item if the variable $x$ has type $\tau$, and the term $t$ has type
  $\sigma$, then $\lambda x.t$ has type $\tau\to\sigma$;
\item if the term $s$ has type $\tau\to\sigma$ and $t$ has type
  $\tau$, then $st$ has type $\sigma$.
\end{enumerate*}
We shall write $t:\tau$ or $t^\tau$ to denote that a term $t$ has type
$\tau$. The $\beta$-rule preserves typing; namely, if $t\bto^*s$ and
$t:\tau$, then $s:\tau$.

A reduction strategy is a set of rules specifying how to reduce a
$\lambda$-term. Roughly speaking, given a reducible term $t$, a
reduction strategy is a function that selects the redex (or the
redexes) of $t$ that must (or among which we can choose the redex
to) be reduced at the next step. A reduction strategy defines a
sub-rewriting system of $\beta$-reduction and, in some cases, if some
$\beta$-reducible term $t$ contains no valid redex for the given
reduction strategy, it introduces new normal forms.

\subsection{Head reduction}
\label{ssec:head-reduction}

Let us say that a $\beta$-redex $(\lambda x.t)s$ is in \emph{outermost
  head position}\footnote{Usually this is simply referred to as
  \emph{head position}. In the following we shall however present a
  larger notion of head position, in which a $\beta$-redex may be in
  head position even if it is inside the body of a $\beta$-redex in
  head position. According to such a new notion of head position, the
  redex reduced by the head reduction is the outermost $\beta$-redex
  in head position.} in $v$ when
$v=\lambda y_1.\ldots \lambda y_k.(\lambda x. t)s u_1\ldots u_h$, and
that $v$ head reduces to
$v'=\lambda y_1.\ldots \lambda y_k.t\{s/x\} u_1\ldots u_h$, written
$v \toh v'$, by reducing its outermost head redex. A term $v$ is in
\emph{head normal form} when
$\lambda y_1.\ldots \lambda y_k.x u_1\ldots u_h$, which in general is
not a $\beta$-normal form, since $u_1,\ldots,u_h$ may contain
$\beta$-redexes. Indeed, the $\beta$-normal form of $t$, if it exists,
can be found by head reducing a term $t$ to its head normal form
$\lambda y_1.\ldots \lambda y_k.x u_1\ldots u_h$ (if any) first, and
then by recursively applying the head reduction strategy to every
$u_i$ and to the subterms of their head normal forms. 

\subsection{Head contexts}
\label{ssec:head-contexts}

As usual, a context $C$ is a term with a hole $\buco$ (a sort of dummy
free variable occurring exactly once in the term)
$C ::= \buco \mid \lambda x. C \mid C t \mid t C$.  Given any term
$t$, by $C[t]$ we denote the term obtained by replacing the hole of
the context $C$ with the term $t$, without performing any variable
renaming; therefore, when the hole is under the scope of a
$\lambda$-abstraction binding the variable $x$, any free occurrence of
$x$ in $t$ is captured in $C[t]$, and becomes bound.

\begin{definition}[H-context, head variable]
  \label{def:H-context}
  A \emph{head context}, or \emph{H-context}, is a context whose hole
  appears in head position. More precisely, H-contexts are defined by
  the following grammar 
  \begin{equation*}
    H ::= \buco \mid \lambda x. H \mid H t.  
  \end{equation*}
  A head context of a term $t$ is any H-context $H$ s.t.\ $t=H[s]$,
  for some term $s$, that we shall say to be in head position in
  $t$. In particular, for every $\lambda$-term $t$, there is a unique
  head context $H$ of $t$ (the \emph{maximal head context} of $t$) and
  a unique variable $x=\hv(t)$ (the \emph{head variable} of $t$) s.t.\
  $t=H[x]$.
\end{definition}

\subsection{Spine}
\label{ssec:spine}

A spine $\lambda$-abstraction/application of a term $t$ is any
$\lambda$-abstraction/application in head position in $t$. The
\emph{spine} of  $t=H[x]$, and of its head context $H$, is the
sequence of its spine $\lambda$-abstractions/applications ordered from
the head variable of $t$ (the hole of $H$) to its root.  A variable
$x$ bound by a spine abstraction is a \emph{spine variable}, while the
right subterm of a spine application is a \emph{spine argument} of
$t$. By $\SV(t)$ and $\SV(H)$ we denote the set of the spine variables
of a term $t$ and of a H-context $H$, respectively.

A H-context $H_\lambda$ is a \emph{$\lambda$-context} if its spine is
formed of $\lambda$-abstractions only (equivalently, $H_\lambda$ has
no spine arguments). A H-context $H_@$ is a \emph{$@$-context} if its
spine is formed of applications only (equivalently, $H_@$ has no spine
variables).

\section{$\beta$-reduction at a distance}
\label{sec:beta-red-at-dist}

\subsection{Environment contexts}
\label{ssec:environment-contexts}

\begin{definition}[E-context]
  \label{def-E-context}
  An \emph{environment context}, or \emph{E-context}, is a particular
  H-context in which spine $\lambda$-abstractions and spine
  applications are balanced.  E-contexts are defined by the
  grammar
 \begin{equation*}
   E_{1,2}::= \buco \mid E_1[\lambda x.E_2]t
 \end{equation*}
\end{definition}
An E-context $E$ contains an equal number $\nprim E$ of spine variables
and of spine arguments. For every E-context $E\neq \buco$, there is a
unique pair $(x,t)$ s.t.\ $E=E_1[\lambda x.E_2]t$, for some pair of
E-contexts $E_1, E_2$.  Therefore, every E-context defines a unique
bijection between its spine variables and its spine arguments. Such a
correspondence can be formalised in terms of environments.
An \emph{environment} $\eta=t_1/x_1, \ldots , t_k/x_k$ is an ordered
sequence of variable substitutions $t_i/x_i$ (where $t_i$ is a term
replacing the variable $x_i$). Given an environment $\eta$, we define
$t\{\eta\}=t\{t_1/x_1, \ldots , t_k/x_k\} =
t\{t_1/x_1\}\ldots\{t_k/x_k\}$.

\begin{definition}
  \label{def:eta-env}
  The environment $\eta(E)$ associated to an E-context is inductively
  defined by 
  \begin{equation*}
    \eta(\buco) = \epsilon
    \qquad\text{ and }\qquad
    \eta(E_1[\lambda x.E_2]t) = \eta(E_2),t/x,\eta(E_1)
  \end{equation*}
\end{definition}

According to the above definition, every pair of matching spine
argument/variable corresponds to a substitution $t/x$ in $\eta(E)$.
We remark that the order of the substitutions in an environment is
relevant, since for $i < j$, the occurrences of $x_j$ in the term
$t_i$ are replaced by the term $t_j$, while this is not the case for
any occurrence of $x_j$ in a term $t_k$ with $k\geq j$.  In
particular, the order of the spine variables in $\eta(E)$ corresponds
to the order in which they appear in $E$, assuming to move from the
inner head position to the root. In other words, $x$ precedes $y$ in
$\eta(E)$ iff the binder of $x$ is in the scope of the binder of $y$.

\begin{lemma}\label{lemma:E-context-beta}
  Let $E$ be an E-context.   For every $\lambda$-term t,
  $E[t] \bto^* t\{\eta(E)\}$.
\end{lemma}

\subsection{Primary redexes and $\beta$-contraction at a distance}
\label{sssec:primary-redexes}

Any pair of matching spine argument/variable in an E-environment is as a
sort of redex at a distance.

\begin{definition}[Primary $\beta$-redex]
  \label{def:primary-redex}
  A \emph{$\beta$-redex at a distance} is a term $E[\lambda x.t]s$, where
  $E$ is an E-context. A \emph{primary $\beta$-redex} is a
  $\beta$-redex at a distance occurring in a head position.
\end{definition}

As a particular case, for $E=\buco$, any $\beta$-redex is a
$\beta$-redex at a distance.  $\beta$-redexes at a distance can be
reduced as usual $\beta$-redexes, by defining the following
generalisation at a distance of the $\beta$-rule
\begin{equation*}
  E[\lambda x.t]s \btod E[t\{s/x\}]
\end{equation*}
and by taking the $\beta_d$-reduction as the closure by contexts of
the above rule. The H-context $E[\lambda x . \buco]s$ of a
$\beta$-redex at a distance is an E-context. Then, every pair $t/x$ of
matching spine argument/variable of an $E$-context (and therefore
every substitution in $\eta(E)$) forms a primary redex. As a
consequence, it is readily seen that $E[t] \btod^* t\{\eta(E)\}$ for
every E-context $E$ and every term $t$. More generally,
$\beta$-reduction at a distance is sound w.r.t.\ the usual
$\beta$-reduction.

\begin{proposition}\label{prop:betad-beta}
  Let $t \btod^* s$, then $t \beq s$. Moreover, $s$ is a normal form
  for $\btod$ iff it is a $\beta$-normal form.
\end{proposition}

\section{Spine permutation equivalence of $\lambda$-terms} 
\label{sec:spine-perm-equiv}

The \emph{head canonical E-contexts} are a particular case of
E-contexts in which every redex at a distance is also a
$\beta$-redex. Head canonical E-contexts are defined by the grammar
$E_c ::= \buco \mid (\lambda x.E_c)t$, and any head canonical
E-context has the shape
\begin{math}
  (\lambda x_n. \ldots (\lambda x_2. (\lambda x_1. \buco) t_1)
  t_2 \ldots) t_n.
\end{math}
An environment $\eta$ can be seen as the explicit representation of
a head canonical E-context $\mathcal{E}(\eta)$ in which the order of the
$\beta$-redexes along the spine is the inverse of the substitution pairs
in the environment
\begin{equation*}
  t_1/x_1,t_2/x_2,\ldots,t_n/x_n 
  \qquad\stackrel{\mathcal{E}}{\longmapsto}\qquad 
  (\lambda x_n. \ldots (\lambda x_2. (\lambda x_1. \buco) t_1) t_2 \ldots)  t_n
\end{equation*}  
Which corresponds to the inductive definition
$\mathcal{E}(\epsilon) = \buco$, and 
$\mathcal{E}(t/x,\eta) = \mathcal{E}(\eta)[(\lambda x.\buco) t]$.

\subsection{Surface E-equivalence} 
\label{ssec:surf-e-equivalence}

By Lemma~\ref{lemma:E-context-beta}, we have that $E_1\beq E_2$, for
every pair of E-contexts $E_1$ and $E_2$ s.t.\
$\eta(E_1)=\eta(E_2)$. We can then define the following equivalence.

\begin{definition}[Surface E-equivalence on E-contexts]
  \label{def:surf-E-eq-contexts}
  The \emph{surface E-equivalence on E-contexts} is the least 
  equivalence $\esim$ defined by
  \begin{align*}
    E_1[\lambda x.E_2]t & \esim E_1[(\lambda x.E_2)t] 
                          && \mbox{if } \FV(t) \cap \SV(E_1)=\emptyset
    \\
    E_1[E_2] & \esim E_1'[E_2'] && \mbox{if } E_i \esim E_i' \text{ for } i=1,2
  \end{align*}
\end{definition}

Such an equivalence captures exactly the equivalence classes of
E-contexts $\{E \mid \mathcal{E}(\eta(E)) = E_c\}$, where $E_c$ is
head canonical, as formally stated by the following lemma.

\begin{lemma}\label{lem:E-context-un-can}
  For every E-context $E$, there is a unique canonical E-context
  $E_c\esim E$, which is also the unique normal form of the
  terminating rewriting system $\eto$ obtained by orienting the
  $E$-equivalence rules of Definition~\ref{def:surf-E-eq-contexts}
  from the left to the right
  \begin{align*}
    E_1[\lambda x.E_2]t & \eto E_1[(\lambda x.E_2)t] 
                          && \text{if } \FV(t) \cap \SV(E_1)=\emptyset
    \\
    E_1[E_2] & \eto E_1'[E_2'] && \text{if } E_i \eto E_i' \text{ and } E_j = E_j',
                                   \text{ with } i,j\in\{1,2\}, \text{ and } i\neq j
  \end{align*}  
  Moreover, $E_c=\mathcal{E}(\eta(E))$, and therefore $E\esim E'$ iff
  $\eta(E)=\eta(E')$.
\end{lemma}

\begin{example}
  \label{ex:E-canon}
  Let $E=E_1[\lambda x.E_2]t$ with $E_1=(\lambda y.\buco)s$ and
  $E_2=\buco$. The E-context
  $E_c = (\lambda y. (\lambda x. \buco)t)s$ is the unique canonical
  E-context $\esim$-equivalent to
  $E = (\lambda y. \lambda x. \buco)s t$.
\end{example}

\subsection{Canonical $\lambda$-terms}
\label{ssec:canon-lambda-terms}

The E-equivalence can be extended to terms. In the corresponding head
canonical forms, along the spine, one finds first all the unmatched
spine abstractions, then the E-context formed of the primary redexes,
and finally the unmatched spine arguments.

\begin{definition}[Surface E-equivalence on terms]
  \label{def:surf-E-eq-terms}
  The \emph{surface E-equivalence on terms} is the least equivalence
  defined by the E-equivalence rules on E-contexts of
  Definition~\ref{def:surf-E-eq-contexts}, plus
  \begin{align*}
    H_\lambda[E[\lambda x. s]] & \esim H_\lambda[\lambda x. E[s]]
    && \text{if } x \not\in FV(E)                             
    \\
    H_\lambda[E[s] t] & \esim H_\lambda[E[s t]]
    && \text{if } FV(t) \cap \SV(E)=\emptyset
    \\
    H_\lambda[E_1[s]] & \esim H_\lambda[E_2[s]]
    && \text{if } E_1 \esim E_2
  \end{align*}
  where $H_\lambda$ is a $\lambda$-context, and
  $E,E_1,E_2$ are E-contexts. The equivalence naturally extends to
  H-contexts, by replacing $\buco$ for $s$ in the above equations.
\end{definition}

\begin{definition}[head canonical $\lambda$-term]
  Let us say that $H$ is a \emph{head canonical H-context} when
  \begin{equation*}
    H=H_\lambda[E_c[H_@]]
  \end{equation*}
  where $H_\lambda$ is a $\lambda$-context, $H_@$ is an $@$-context,
  and $E_c$ is a head canonical E-context. The spine
  $\lambda$-abstractions of $H_\lambda$ are the \emph{head
    $\lambda$-abstractions} of $H$, while the spine arguments of $H_@$
  are the \emph{head arguments} of $H$.
  The $\lambda$-term $t$ is \emph{head canonical} when its maximal
  head context is head canonical.
\end{definition}

Summing up, any head canonical $\lambda$-term $t$
has the shape
\begin{equation*}
  \begin{aligned}
    t  & =\lambda x_1. \ldots \lambda x_n. E_c[z\, t_1 \ldots t_m] 
    \\ & =
    \lambda x_1. \ldots \lambda x_n. (\lambda y_1 . (\ldots (\lambda y_p. z\, t_1 \ldots t_m) s_p) \ldots) s_1)
  \end{aligned}
\end{equation*}
and we can define $\nlam t = n$, $\narg t = m$, $\eta(t)=E_c$, and
$\nprim t = \nprim E_c = p$.

Every H-context $H$, and then every $\lambda$-term $t=H[x]$, has a
unique E-equivalent head canonical form $H_c$, or $H_c[x]$ for
terms. Moreover, as shown by Theorem~\ref{thm:un-head-can} below,
$H_c$ preserves the same relative positions of unmatched spine
$\lambda$-abstractions, unmatched spine arguments, and primary redexes
of $H$. (A spine $\lambda$-abstraction/argument is unmatched when it
is not involved in a primary redex.)  More precisely, the $i$-th head
$\lambda$-abstraction of $H_c$ is the $i$-th unmatched
$\lambda$-abstraction on the spine of $H$, the $i$-th head argument of
the head canonical form is the $i$-th unmatched spine argument on the
spine of $H$, the $i$-th primary redex of $H_c$ is the $i$-th primary
redex on the spine of $H$.

\begin{theorem}\label{thm:un-head-can}
  For any H-context $H$, there is a unique head canonical
  context $H_c \esim H$. More precisely, 
  \begin{enumerate}
  \item\label{item:lem:un-head-can:spine} %
    for every H-context $H$,  there is a unique sequence of spine
    variables $x_1, \ldots x_n$, a unique sequence of spine arguments
    $t_1,\ldots,t_m$, and a unique sequence of E-contexts
    $E_0,E_1,\ldots,E_{n+m}$ s.t.\
    \begin{align*}
      H & = E_0[H_1] \\
      H_i &= \lambda x_i. E_i[H_{i+1}] &\qquad \mbox{for $1\leq i\leq n$} \\
      H_{n+j} &= E_{n+j}[H_{n+j+1}]t_{m-j+1} &\qquad \mbox{for $1\leq j\leq m$} \\
      H_{n+m+1} &= \buco
    \end{align*}
    that is
    \begin{equation*}
      H = E_0[\lambda x_1.E_1[\lambda x_2. E_2[\ldots [\lambda x_n.E_{n}
      [E_{n+1}[\ldots[E_{n+m-1} [E_{n+m} t_1]t_2]\ldots]t_m]]\ldots]]]
    \end{equation*}
  \item\label{item:lem:un-head-can:can-form} %
    there is a unique head canonical context
    $H_c \esim H$, and $H_c = H_\lambda[E[H_@]]$ is equal to
    \begin{align*}
      H_\lambda &= \lambda x_1.\ldots \lambda x_n. \buco\\
      H_@ &= \buco t_1 \ldots t_m \\
      E_c &= \widetilde{E}_0[\widetilde{E}_1[\ldots[\widetilde{E}_{n+m}]]]
    \end{align*}
    that is
    \begin{equation*}
      H \esim H_c = \lambda x_1. \ldots \lambda x_n. E_c[\buco\, t_1 \ldots t_m]
    \end{equation*}
    where $\widetilde{E}_i=\mathcal{E}(\eta(E_i))\esim E_i$ is the 
    unique head  canonical E-context equivalent to $E_i$;
  \item the canonical context $H_c$ of $H$ is the unique normal form
    of the rewriting system $\eto$ obtained by orienting from the left to
    the right the surface E-equivalences on terms of
    Definition~\ref{def:surf-E-eq-terms}. Namely,
  \begin{align*}
    H_\lambda[E[\lambda x. s]] & \eto H_\lambda[\lambda x. E[s]]
    && \text{if } x \not\in FV(E)                             
    \\
    H_\lambda[E[s] t] & \eto H_\lambda[E[s t]]
    && \text{if } FV(t) \cap \SV(E)=\emptyset
    \\
    H_\lambda[E_1[s]] & \eto H_\lambda[E_2[s]]
    && \text{if } E_1 \eto E_2
  \end{align*}
  plus the rules for E-contexts in
  Lemma~\ref{lem:E-context-un-can}.
  \end{enumerate}
\end{theorem}

\subsection{E-equivalence}
\label{ssec:e-equivalence}

The surface E-equivalence permutes the arguments on the spine of a
term without modifying them. The E-equivalence is obtained by
recursively applying the surface E-equivalence to spine arguments
too. If we denote by $\ar(t,i)$ the $i$-th head spine
argument of the term $t$ (which corresponds to the $i$-th spine
argument in the head $@$-context of its head canonical form) and by
$\ar(t,-i)$ the spine argument of the $i$-th primary redex of $t$
(which corresponds to the $i$-th spine argument in the head canonical
E-context $E_c$ of the head canonical form of $t$), we define $\econg$
as the least equivalence s.t.\ $t_1 \econg t_2$ if $t_1 \esim t_2$,
and $\ar(t_1,i) \econg \ar(t_2,i)$, for
$1 \leq i \leq \narg t_1 = \narg t_2$ or
$-\nprim t_2 = -\nprim t_1 \leq i \leq -1$.

\subsection{$\sigma$-equivalence}
\label{ssec:sigma-equivalence}

The \emph{$\sigma$-equivalence} \cite{Reg:EquivLamTermes:94} is the
least congruence induced by
\begin{align*}
  ((\lambda x. u) v) w & \sigcong (\lambda x. u w)v
  && \text{with } x\not\in \FV(w)
  \\
  (\lambda x.\lambda y. u)v & \sigcong \lambda y.(\lambda x. u)v
  && \text{with } y \not\in \FV(v)
  \end{align*}

The rewriting system obtained by orienting the latter
$\sigma$-equivalences from the left to the right is terminating---its head
canonical forms are the same already defined for the
E-equivalence---but is not confluent.  Indeed, the 
$\sigma$-equivalence contains the E-equivalence, but it equates head
canonical forms $E_1$ and $E_2$ s.t.\ the environments $\eta(E_1)$ and
$\eta(E_2)$ are equivalent modulo the following permutation rule
$t_1/x_1,t_2/x_2\sim t_2/x_2,t_1/x_1$ if $x_1\not\in\FV(t_2)$ and
$x_2\not\in\FV(t_1)$.

\begin{example}
  \label{ex:sig-can}
  Let us take the $\lambda$-term
  $u = E[v] = (\lambda y. \lambda x. v)s t$, where $E$ is the
  $E$-context of Example~\ref{ex:E-canon}. Its unique head E-canonical
  form is $(\lambda y. (\lambda x. v) t)s$, which can be also obtained
  by applying the first $\sigma$-rule. However, since by applying the second
  $\sigma$-rule, $u \sigto (\lambda x. (\lambda y. v) s)t$
  too, the $\lambda$-term $u$ has two $\sigma$-equivalent canonical forms.
\end{example}
 
Summing up, the E-equivalence is a variant of the $\sigma$-equivalence
which equates less terms then the latter one.  The definition of the
$\sigma$-equivalence is simpler and more elegant, and has a direct and
nice interpretation in terms of linear logic proof nets. However, the
better rewriting properties of the E-equivalence---canonical form
uniqueness and preservation of primary redexes relative
positions---makes it more suitable for a finer analysis of reduction
machines requiring a reduction at a distance based on
$\sigma$-equivalence, as for instance the PAM. The
$\sigma$-equivalence can be recovered from the E-equivalence by adding
the following permutation equivalence of primary redexes
\begin{equation*}
  E_1[(\lambda x_1 . (\lambda x_2. E_2) t_2) t_1] \sim E_1[(\lambda
  x_2. (\lambda x_1. E_2) t_1) t_2]
\end{equation*}
if $x_1\not\in\FV(t_2)$ and $x_2\not\in\FV(t_1)$, to the E-equivalence
of E-contexts.

\section{Linear head reduction}
\label{sec:line-head-reduct}

\subsection{Linear reduction}
\label{ssec:linear-reduction}

Let $(\lambda x.t)s$ be a redex s.t.\ the term $t$ contains at least
one occurrence of $x$. For any occurrence of $x$ in $t$, we can take the
context $C$ obtained by replacing such an occurrence of
$x$ with $\buco$.  The following reduction rule
\begin{math}
  (\lambda x. C[x]) s \ltob (\lambda x. C[s'])s,
\end{math}
where $s'$ is a fresh copy of $s$, is a linearised variant of the
usual $\beta$-rule in which, instead of removing the redex after
replacing all the occurrences of the bound variable $x$, the redex is
kept and only one occurrence of $x$ is replaced by a fresh copy of the
argument $s$. Such a linear $\beta$-reduction can be extended
to be applied at a distance too. We obtain then the \emph{linear
  reduction rule} (at a distance)
\begin{equation*}
  E[\lambda x.C[x]]s \lto E[\lambda x.C[s']]s
\end{equation*}
where $s'$ is a fresh copy of $s$.  When the term $t$ in
$E[\lambda x.t]s$ does not contain any occurrence of $x$, we can
instead take the following \emph{garbage rule} (which is just a
degenerated case of $\beta$-reduction at a distance)
\begin{equation*}
  E[\lambda x. t] s \gto E[t] \qquad \text{if } x\not\in\FV(t)
\end{equation*}

Given a $\beta$-redex (at a distance), by iterating the linear
$\beta$-reduction (at a distance), we can eventually obtain a redex
(at a distance) to which apply the \emph{garbage rule}. Therefore,
$\beta$-reduction (at a distance) can be simulated by a sort of affine
reduction $\ato$ which is the union of linear and 
garbage reduction.

\begin{proposition}\label{prop:aff-red}
  Let $\ato = \lto \cup \gto$.
  \begin{enumerate}
  \item If $t \bto^* s$, then $t \ato^* s$. Moreover, there is $s'$ s.t.\ 
    $t \lto^* s' \gto^* s$.
  \item If $t \ato^* u$, then $u \beq t$. Therefore, there is $t \bto^* s$ s.t.\
    $u \ato^* s$.
  \end{enumerate}
\end{proposition}

As a consequence of the above proposition, a term has a normal form
for $\ato$ iff it has a $\beta$-normal form; moreover, the two normal
forms coincide. We also remark the second part of the first item of
Proposition~\ref{prop:aff-red}. This is a particular case of a more
general property stating that garbage reductions can be always
postponed; that is, for every
$t \ato^* s$, there is $s'$ s.t.\ $t \lto^* s' \gto^* s$.

\subsection{Linear head $\beta$-rule}
\label{ssec:line-head-contr}

A particular case of linear reduction arises when the occurrence to be replaced 
is the head variable.

\begin{definition}[Linear head reduction]
  The \emph{linear head reduction} is the least reduction which
  contains the linear head $\beta$-rule
  \begin{equation*}
    E[\lambda x.H[x]]s\ltoh E[\lambda x.H[s']]s
  \end{equation*}
  where $s'$ is a fresh copy of $s$, and which is closed by head contexts.
\end{definition}

Linear head reduction is strongly related to head $\beta$-reduction,
as shown by the following statements.

\begin{proposition}\label{prop:hnf-lhnf}
  Let $t \ltoh^* s$. There is $t \toh^* s'$ s.t.\ $s \toh^* s'$.
\end{proposition}

\begin{corollary}\label{cor:lhnf-hnf}
  A term $t$ has a linear head normal form iff it has a head normal
  form. Moreover, let $s$ be the linear head normal form of $t$.
  \begin{enumerate}
  \item The head normal form of $s$ is obtained by
    $\beta$-reducing all the primary redexes in $s$.
  \item The head normal form of $s$ is the head normal form of $t$,
    indeed.
  \end{enumerate}
\end{corollary}

\section{Strong normalisation}
\label{sec:strong-normalisation}

All the rewriting systems defined above are strong normalising on
simply typed $\lambda$-terms.  The proof of strong normalisation is
however not at all evident. In fact, since linear reduction does not
erase the reducing redex---it just replaces the occurrence of a
variable by a (larger) $\lambda$-term---the size of the reducing term
increases at each step.  Accattoli~\cite{accattoli:LIPIcs:2013:4051},
in its analysis of proof nets linear reduction, proved strong
normalisation by applying reducibility candidates. Here, we show that,
surprisingly, the proof of strong normalisation of linear reduction is
simpler then one might have thought, as it can be easily obtained by a
trivial tuning of the proof of strong normalisation originally
proposed by Gandy for $\beta$-reduction~\cite{Gan:ProofsSN:80}. In
Gandy's proof, each type $\tau$ is interpreted as a well-founded
ordered set $\sfun{\tau}$. In particular, any functional type
$\tau\to\sigma$ is mapped into a set of increasing functions from
$\sfun{\tau}$ to $\sfun{\sigma}$. A measure is then associated to
every term by interpreting any $t:\tau$ as an element
$\sfun{t}\in\sfun{\tau}$. Strong normalisation is a consequence of the
fact that any $\beta$-reduction $t \bto s$ sends $\sfun{t}$ to a lower
element $\sfun{s}$.

The original measure defined for the analysis of $\beta$-reduction
does not directly work for the case of linear reduction, since such a
measure does not change along linear reduction (i.e.,
$\sfun{t} = \sfun{s}$, when $t \lto s$). Indeed, Gandy's
measure just counts the number of $\lambda$-abstractions erased along
a $\beta$-reduction. However, by taking the successor of the
usual interpretation of a variable occurrence, one obtains a new
measure which counts the number of variable occurrences replaced by
some $\lambda$-term. Such a new measure decreases along linear
reduction, and allows to prove at the same time the strong
normalisation of all the rewriting systems described in the present papers.

In the following, we shall follow the presentation of
Gandy's proof given by Miquel~\cite{Miq:CombProof}.  Let us interpret
the base type $o$ as the strict partial order $(\NN,<)$, and every
functional type $\tau\to\sigma$ as the strict partial order of the
increasing functions from the interpretation of $\tau$ to the
interpretation of $\sigma$. Formally, for every type $\tau$, let us
inductively define $(\dom{\tau}, \spo_\tau)$ by
\begin{equation*}
  \dom{\tau \to \sigma}  = \left \{
    f \in \dom{\tau} \to \dom{\sigma} \mid \forall v, w\in\dom{\tau} : v \spo_\tau w \imp f(v) \spo_\sigma f(w)
  \right \}
\end{equation*}
\begin{equation*}
 \forall f,g \in \dom{\tau\to\sigma} \quad : \quad 
  f \spo_{\tau\to\sigma} g \quad \text{ iff }\quad \forall v\in\dom{\tau}\ :\ 
   f(v) \spo_{\sigma} g(v)
\end{equation*}
with $\dom{o}=\NN$ and ${\spo_o}={<}$. We define then the binary operation
$+_\tau:\dom{\tau}\times\NN \to \dom{\tau}$ as
\begin{equation*}
  n +_o k = n + k \qquad\qquad f +_{\dom{\tau\to\sigma}} k =
  (\lambda v \in \dom{\tau}. f(v) +_{\sigma} k)
\end{equation*}
for $n,k\in\NN$ and $f\in\dom{\tau\to\sigma}$. It is readily seen that
$v +_\tau 0 =v$, that $(v+_{\tau}k)+h=v+_{\tau}+(k+h)$, and that
$k < h$ implies $v +_\tau k \spo v +_\tau h$,
for every $v\in\dom{\tau}$ and $k,h\in\NN$.

For every type $\tau$, let us define $\tau_*\in\dom{\tau}$ and
$\tau^*:\dom{\tau}\to\NN$ by
\begin{align*}
  o_* &\ =\ 0 
  & (\tau \to \sigma)_* 
  &\ =\ (\lambda v \in \dom{\tau} . \sigma_* 
    +_\sigma \tau^*(v)) 
  \\
  o^*(n) &\ =\ n
  & (\tau \to \sigma)^*(f)
  &\ =\ \sigma^* (f(\tau_*)) 
\end{align*}
for $n\in\NN$ and $f\in\dom{\tau\to\sigma}$. By induction, we can see
that $\tau^*$ is increasing (that is, $\tau^*(v) < \tau^*(w)$, for all
$v,w\in\dom{\tau}$ s.t.\ $v\spo_\tau w$). 

A \emph{valuation} is a function $\phi$ associating an element of
$\dom{\tau}$ to every variable $x:\tau$. Given a valuation $\phi$, a
variable $x:\tau$, and a value $v\in\dom{\tau}$, we shall denote by
$\phi[x \mapsto v]$ a new valuation s.t.\ $\phi[x \mapsto v](x) = v$,
and $\phi[x \mapsto v](y) = \phi(y)$, when $y\neq x$.

Given a valuation $\phi$, any typed $\lambda$-term $t^\tau$ can be
interpreted as an element $\sfun{t}_\phi\in\dom{\tau}$ by application
of the following inductive definition
\begin{align*}
  \sfun{x:\tau}_{\phi} &\ = \ \phi(x) +_\tau 1 \\
  \sfun{\lambda x. t:\tau\to\sigma}_\phi &\ = 
                                 \ \lambda v\in\dom{\tau}. \sfun{t}_{\phi[x\mapsto v]}
                                 +_\sigma (\tau^*(v) + 1) \\
  \sfun{s^{\tau\to\sigma}t:\sigma}_\phi &\ = \ \sfun{s:\tau\to\sigma}_\phi\sfun{t:\sigma}_\phi
\end{align*}
For every valuation $\phi$, we can also define the measure
$\mu_\phi:\Lambdat\to\NN$, by
$\mu_\phi(t^\tau) = \tau^*\sfun{t}_{\phi}$.

\begin{remark}
  The only difference w.r.t.\ the usual interpretation used in the
  proof of strong normalisation of $\beta$-reduction is the
  interpretation of variables. Indeed, one usually takes
  $\sfun{x:\tau}_{\phi} = \phi(x)$ (see~\cite{Miq:CombProof}). With
  this choice, however, we would get $\sfun{t}=\sfun{s}$ when
  $t \lto s$.
\end{remark}

\begin{lemma}
  For every valuation $\phi$, every $C[x^\tau]:\sigma$, and every
  $t:\tau$, we have that
  \begin{enumerate}
  \item $\sfun{C[x^\tau]}_{\phi[x\mapsto \sfun{t}_\phi]} \spo_\sigma \sfun{C[t]}_\phi$;
  \item if $t \ato s$, then $\sfun{s}_\phi \spo_\sigma \sfun{t}_\phi$ and 
    $\mu_\phi(s) < \mu_\phi(t)$.
  \end{enumerate}
\end{lemma}

By the previous lemma, and the fact that there is at least a valuation
(for instance, the valuation $\phi_0$ defined by
$\phi_0(x^\tau)=\tau_*$), we can eventually get the strong normalisation
result.

\begin{theorem}
  The rewriting systems $\ato$, $\lto$, $\btod$, $\bto$, $\toh$, and
  $\ltoh$ are strongly normalising.
\end{theorem}

\section{Conclusions}
\label{sec:conclusions}

In the paper we have analysed linear $\beta$-reduction in terms of a
notion of evaluation context, and we have seen how a simple adaptation
of the semantical proof of strong normalisation for the simply typed
$\lambda$-calculus allows to prove the same result for the linear
case.  The proof is surprisingly simple and its idea might be adapted
to prove strong normalisations of other $\lambda$-calculi in which the
$\beta$-rule is decomposed in more elementary steps, as for instance
in the case of explicit substitution $\lambda$-calculi.



\bibliographystyle{eptcs}
\bibliography{main}

\newpage


\end{document}